\documentclass{article}
\usepackage[utf8]{inputenc}
\usepackage{graphicx} 

\usepackage{fancyhdr}

\usepackage{mwe} 

\usepackage{amsmath}
\usepackage{amsthm}
\usepackage{amssymb}
\usepackage{comment}
\usepackage{bbm}
\usepackage{soul, color}
\usepackage{appendix}
\usepackage{tasks}
\usepackage{wrapfig}
\usepackage{appendix}
\usepackage{graphicx}
\usepackage[utf8]{inputenc}
 \usepackage[small]{titlesec} 

\usepackage{hyperref}
\usepackage{lipsum}
\usepackage{footnote}
\usepackage{tablefootnote} 
\makesavenoteenv{table}

\newbox{\bigpicturebox}

\usepackage{array}
\usepackage{makecell}

\usepackage{amsmath}

\numberwithin{equation}{section}

\usepackage[ruled,vlined]{algorithm2e}
\usepackage{subcaption,graphicx}

\usepackage{bm}
\usepackage{graphicx}

\newcommand{\N}{\ensuremath{\mathbb{N}}}
\renewcommand{\S}{\mathcal{S}}
\renewcommand{\L}{\mathcal{L}}

\newcommand{\Z}{\ensuremath{\mathbb{Z}}}
\newcommand{\R}{\ensuremath{\mathbb{R}}}
\newcommand{\C}{\ensuremath{\mathbb{C}}}
\newcommand{\T}{\ensuremath{\mathbb{T}}}

\newcommand{\X}{\ensuremath{\mathcal{X}}}

\newcommand{\x}{\mathbf{x}}
\newcommand{\q}{\mathbf{q}}
\newcommand{\vv}{\mathbf{v}}

\newcommand{\SD}{\Sigma \Delta}

\newcommand{\p}{\mathbf{p}}
\newcommand{\Pp}{\mathcal{P}}
\newcommand{\ee}{\mathbf{e}}
\newcommand{\F}{\mathbf{F}}
\newcommand{\Ind}{\mathbbm{1}}

\newcommand{\lam}{\lambda}

\newcommand{\bb}[1]{\mathbf{#1}}

\newcommand{\im}{\mathrm{i}}
\newcommand{\dx}{\mathrm{d}}
 
\newcommand{\e}{\mathrm{e}}

\newcommand{\B}{\mathcal{B}}

\newcommand{\W}{ \bb W}

\newcommand{\sign}{\mathrm{sign}}

\newcommand{\supp}{\mathrm{supp}}

\newcommand{\lbc}{\left\{}
\newcommand{\rbc}{\right\}}

\newcommand{\nofty}[1]{\left\|{#1}\right\|_{\infty}}

\usepackage{relsize}

\usepackage{lipsum} 
\usepackage[OT2,T1]{fontenc}
\DeclareSymbolFont{cyrletters}{OT2}{wncyr}{m}{n}
\DeclareMathSymbol{\Sha}{\mathalpha}{cyrletters}{"58}
 \theoremstyle{plain}
 \newtheorem{thm}{Theorem}[section]

 \theoremstyle{definition}

\theoremstyle{remark}

\title{The Mathematics of Dots and Pixels: On the Theoretical Foundations of Image Halftoning} 
\author{Felix Krahmer, Anna Veselovska\\[6pt]
\small  Technical University of Munich\\[.25mm] 
{\footnotesize \it Department of  Mathematics
and  Munich Data Science Institute}\\[2.5mm]
\small Munich Center for Machine Learning
}
\date{ }

\begin{document}

\sloppy

\maketitle

\begin{abstract}

The evolution of image halftoning, from its analog roots to contemporary digital methodologies, encapsulates a fascinating journey marked by technological advancements and creative innovations. Yet the theoretical understanding of halftoning is much more recent. 

In this article, we explore various approaches towards shedding light on the design of halftoning approaches and why they work. We discuss both halftoning in a continuous domain and on a pixel grid.

We start by reviewing the mathematical foundation of the so-called electrostatic halftoning method, which departed from the heuristic of considering the back dots of the halftoned image as charged particles attracted by the grey values of the image in combination with mutual repulsion.  Such an attraction-repulsion model can be mathematically represented via an energy  functional in a
reproducing kernel Hilbert space allowing for a rigorous analysis of the resulting optimization problem as well as a convergence analysis in a suitable topology.

A second class of methods that we discuss in detail is the class of error diffusion schemes, arguably among the most popular halftoning techniques due to their ability to work directly on a pixel grid and their ease of application.
The main idea of these schemes is to choose the locations of the black pixels via a recurrence relation designed to agree with the image in terms of the local averages. We discuss some recent mathematical understanding of these methods that is based on a connection to $\SD$  quantizers, a popular class of algorithms for analog-to-digital conversion.





\end{abstract}

\section{Introduction}
A key challenge in image representation is that while images modeled as mathematical functions admit values in a continuous interval, both in screen representations and prints, the intensities are typically restricted to a discrete set. For screen representations, typically 256 intensity values per color channel are admissible; for printing only two -- either ink in the respective color is printed or not in a location.

The key idea allowing to address this challenge is that one can exploit an illusion of the visual system: The human eye acts as a low-pass filter when perceiving visual information from a distance, seamlessly blending fine details and capturing the overall intensity.

\begin{figure}[ht!]
\begin{center}
  \subcaptionbox{ 
  \centering
  \label{fig11:a}}{\includegraphics[width=2.1in]{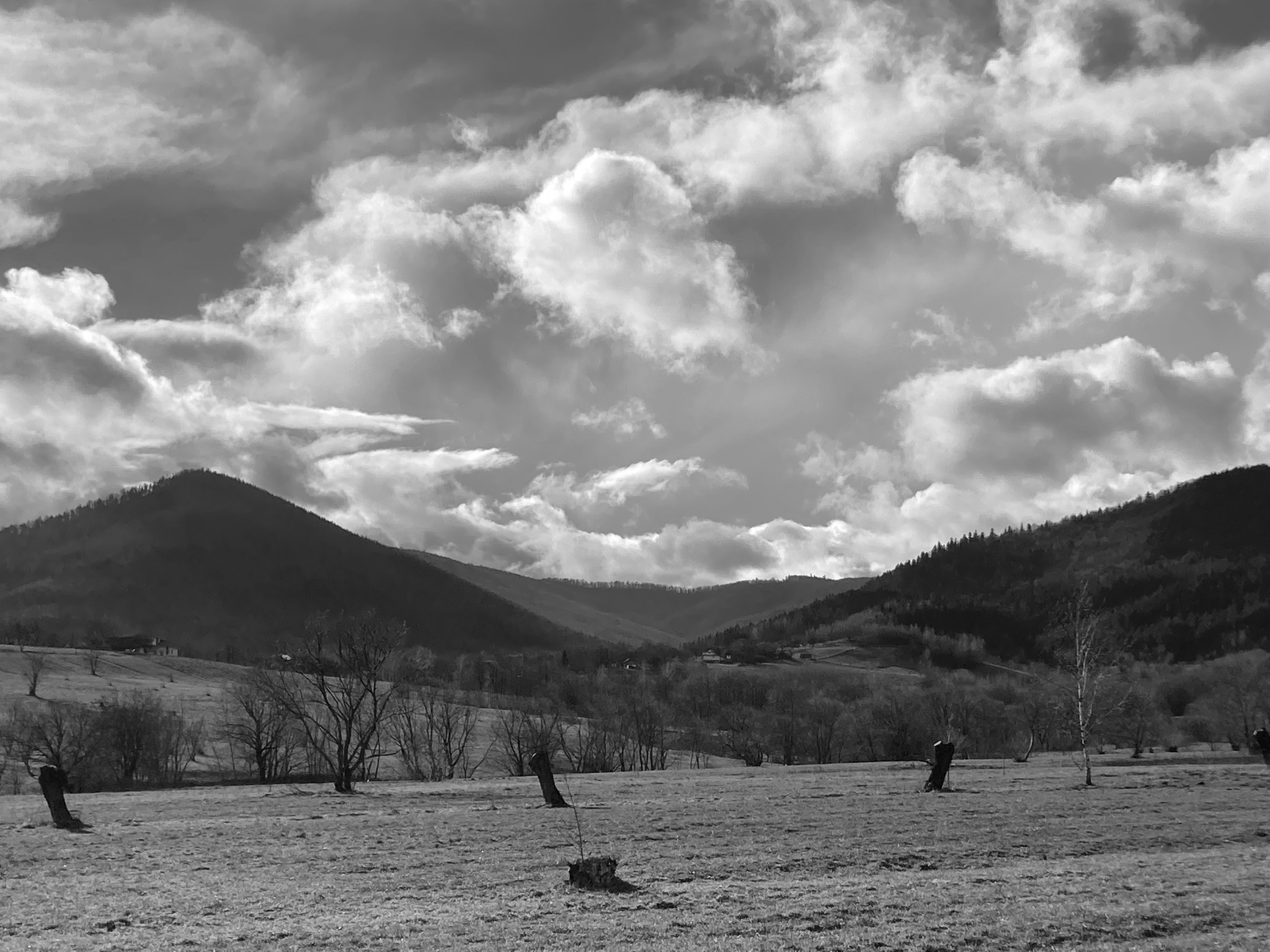}}
 \hspace{10mm}
  \subcaptionbox{  
  \label{fig11:b}}{\includegraphics[width=2.1in]{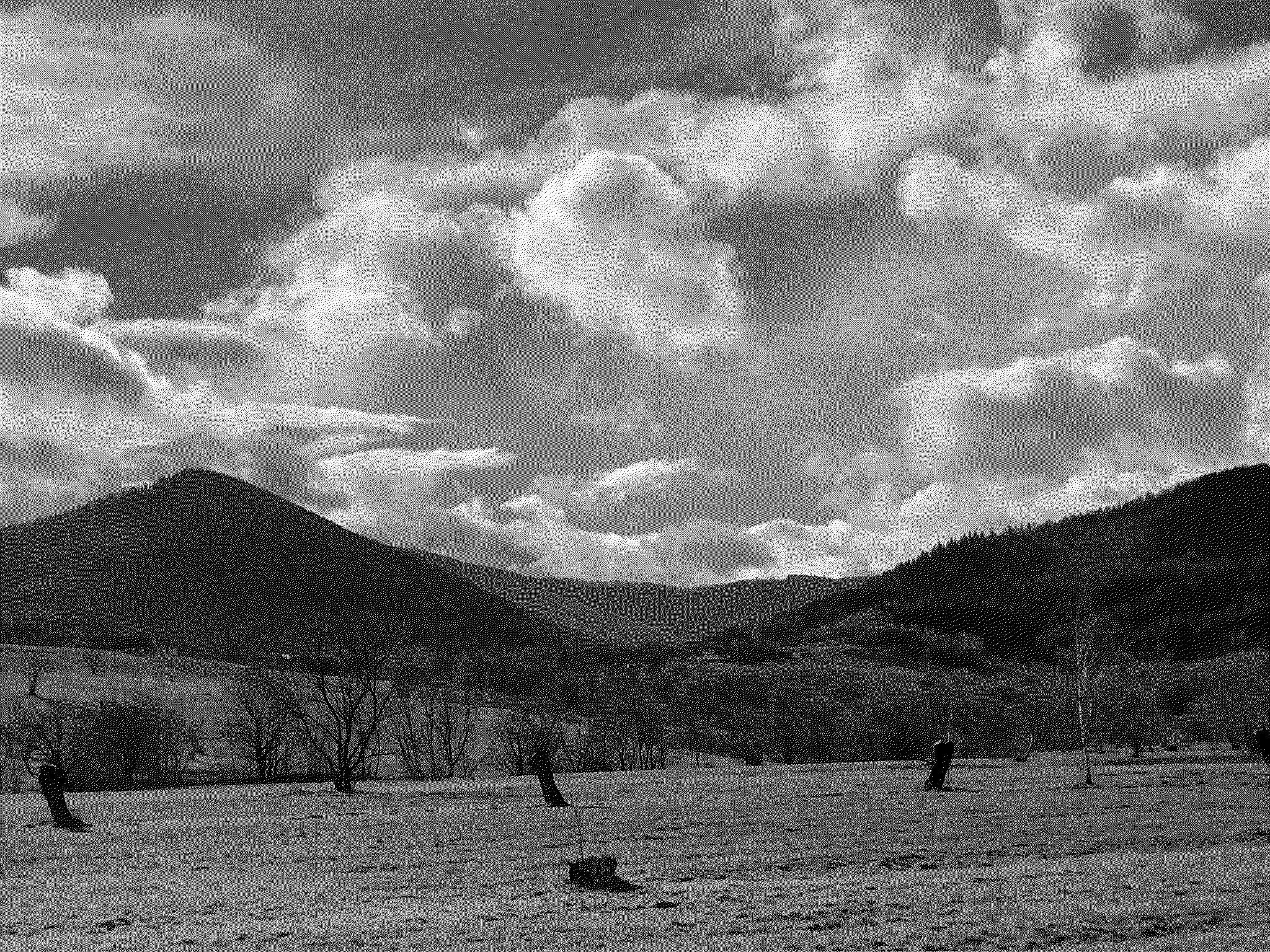}}%
 \caption{Illustration of digital halftoning:
 (a) the original gray-scale image,  (b)~the same image represented by black and white pixels using the Floyd–Steinberg algorithm.}
   \label{fig:cat:1}
   
   \end{center}
\end{figure}  

Building on this observation, the image representation paradigm of halftoning, dating back to 1869, emulates continuous-tone visuals using dots. The mathematical idea behind halftoning is that while the fine details will necessarily be very different, it can be possible to place the dots in such a way that its low-frequency content resembles the image, in this way yielding visual similarity due to the aforementioned illusion.
An illustration of halftoning is provided in Figure~\ref{fig:cat:1}:  the left side showcases a grayscale image, while the right side is meticulously composed using black and white pixels arranged to visually mimic the grayscale image.

\begin{figure}[ht!]
\begin{center}
  \subcaptionbox{ 
  \centering
  \label{fig12:a}}{\includegraphics[width=2in, height=2.4in]{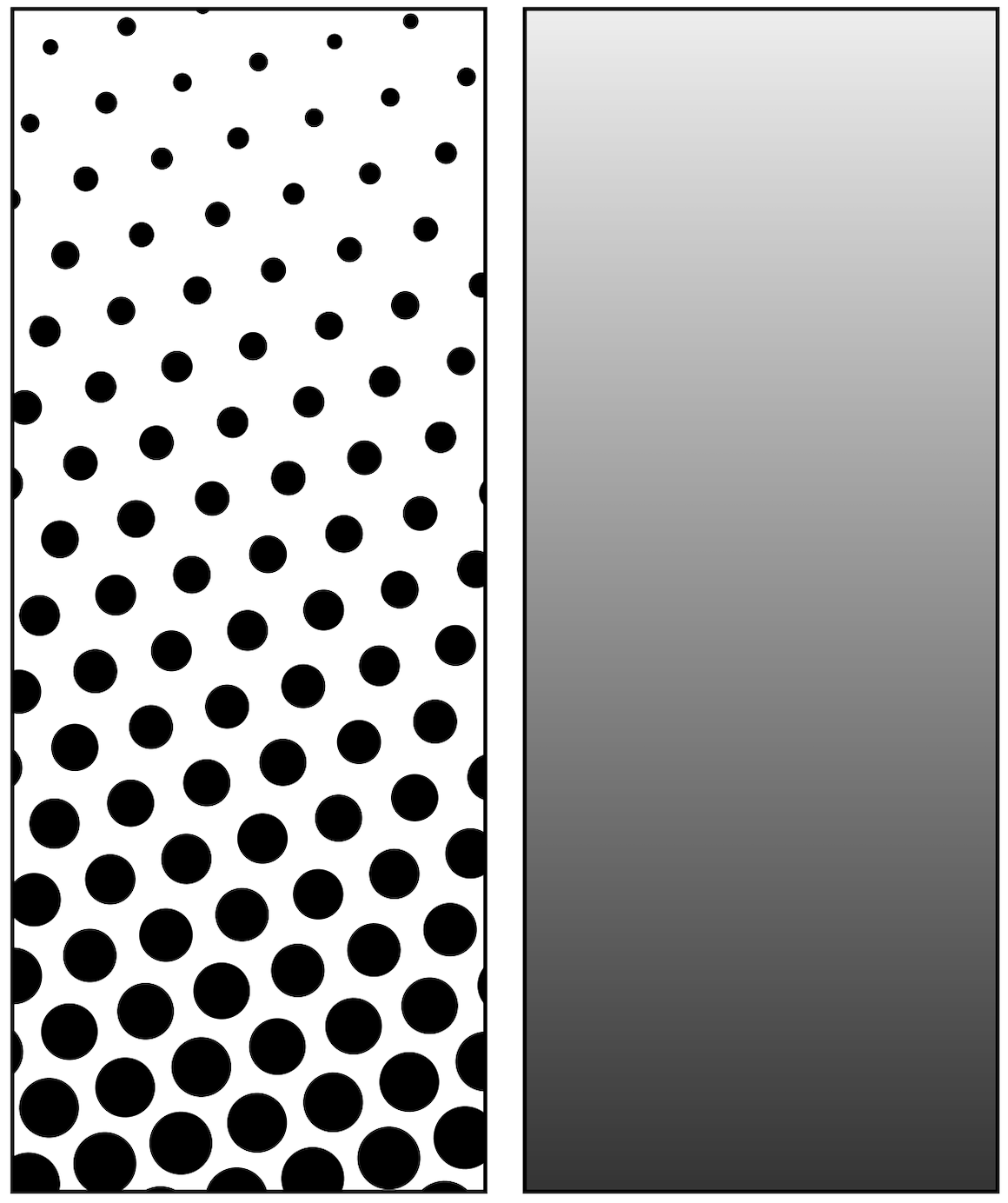}}
 \hspace{10mm}
  \subcaptionbox{  
  \label{fig12:b}}{\includegraphics[width=2.2in]{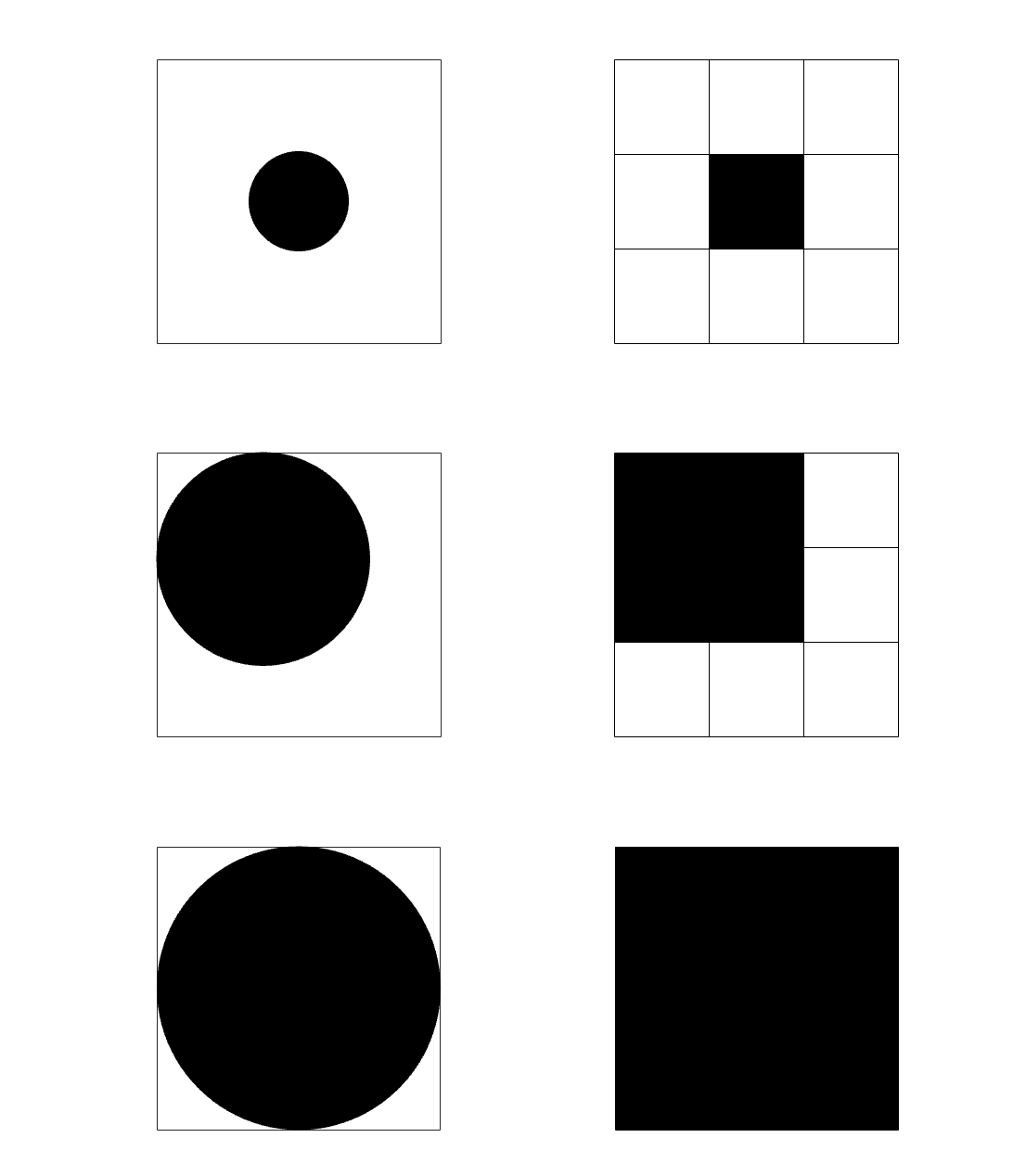}}%
 \caption{
 (a) Illustration of halftoning: gray-scale (on the right) and arrangement of black dots to initiate gray-scale (on the left);  (b) Analog vs. digital halftoning: in the analog halftoning different shades of gray are simulated by black dots of different size, and in the digital one different amount of pixels are turned to black.}
   \label{fig:analog_vs_digital}
   
   \end{center}
\end{figure} 

Halftoning is a well-established area of research in applied imaging science. Its versatility positions it as a valuable tool across a spectrum of applications, showcasing its adaptability and significance in modern visual processing.
Examples include displaying or printing on devices with finite reproduction palettes, e.g. conversion of a 24-bit color image to a three-bit color image and conversion of an 8-bit grayscale image to a binary image \cite{monga2004tone, monga2006design}, tackling sampling issues in rendering \cite{Ostromoukhov2008}, optimizing re-lighting processes \cite{Kollig2003}, influencing object placement strategies, and contributing to artistic non-photorealistic image visualization \cite{Secord2002, Balzer2009}. 

Over the past few decades, halftoning techniques have undergone significant advancements, evolving from basic thresholding and ordered dithering methods approximating color values locally by predefined regular local patterns \cite{Bayer1976} and by more
random patterns as in the method proposed by \cite{Purgathofer1994} to more sophisticated strategies like structure-aware halftoning \cite{Pang2008}. At the same time, most of these methods are heuristic in nature, exhibiting good performance, but without much underlying theory.

Mathematical investigations of halftoning started much later and are still an active area of research. In this survey paper, we will give an overview of the developments as well as some recent advancements.

Starting point of the theoretical underpinning was the groundbreaking idea by Weickert et al.\cite{Weickert2010}~to view black halftoning dots as small charged particles in a 2-D particle system on a bounded domain with attraction and repulsion forces. The setup is continuous in the sense that the particles are not confined to a grid or another structured subset. 
Modeling the repulsion and attraction potentials as a difference of two convex functions allowed employing the theory of convex analysis to obtain a first mathematical analysis of the halftoning process \cite{Teuber2011}.
Subsequent work modeling images as probability measures \cite{Fornasier2016} yields a mathematical analysis for more  general attraction-repulsion functionals. Namely, it has been shown that the resulting halftoned images (in the form of a weighted sum of Dirac measures) 
$\Gamma$-converge to the target energy with respect to the narrow topology on the space of probability measures \cite{Fornasier2016}.  
We will discuss the results of halftoning via attraction-repulsion models in more detail in Section~\ref{sec:rip-att-haftoning} below.

It should be noted that in attraction-repulsion approaches for halftoning, the resulting dot configurations admit general real-valued particle positions -- halftoning with such a setup is often referred to as {\em analog halftoning}. In contrast, many applications impose some underlying structure such as a pixel grid -- this is often referred to as {\em digital halftoning}. The difference between analog and digital halftoning is illustrated in Figure~\ref{fig:analog_vs_digital}.
In \cite{Teuber2011}, the discretization effects resulting from rounding to the nearest grid points have been studied empirically, but to the best of our knowledge, these effects are not covered by the theory yet. At the same time, most approaches that are intrinsically discrete are entirely heuristic in nature and, until recently, had not no theoretical underpinning.

A class of intrinsically discrete methods for digital halftoning that have been particularly popular, partly due to their simplicity, is that of error diffusion methods. The idea is to use a recursive method to achieve a halftoned image with local averages similar to the original image.
A crucial step towards a theoretical foundation was taken in \cite{Knox1992, Kite1997} by observing a deep conceptual analogy between the halftoning problem and the problem of $\SD$ modulation, for which much more theory has been developed. Most of these results on $\SD$ modulation, however, do not directly carry over to the error diffusion problem, as they are either specific to one-dimensional signals (e.g., \cite{Gunturk2003, Daubechies2003, deift2011optimal}) or do not apply to binary scenarios (e.g., \cite{lyu2023sigma, yilmaz2005coarse}). As shown in \cite{Krahmer2023-siam}, however, an appropriate weighted combination of one-dimensional schemes inherits some of the one-dimensional guarantees while at the same time yielding state-of-the-art empirical performance. The mathematical analysis of error diffusion schemes based on $\SD$ modulation is discussed in more detail in Section~\ref{sec:error-diff-halftoning}.

The ideas of image halftoning were also extended to some manifold and graph domains, both in analog \cite{graf2012quadrature, ehler2021curve} and digital scenarios \cite{graf2023one, graf2019higher, krahmer2023quantization}.   Moreover, several research groups made the first steps towards video compression via halftoning techniques \cite{hild19893, gotsman1993halftoning, rehman2010flicker, rehman2011alleviating, schmaltz2012video}. 
In these directions, the mathematical understanding of video digital halftoning and halftoning on manifolds using error diffusion schemes is still a question for future research.


\section{ Attraction-Repulsion Methods for Image Halftoning}\label{sec:rip-att-haftoning}

A starting point for understanding the halftoning problem is that the filtering operation modeling human vision performs a kind of local averaging. That is, the local density of the distribution of dots in the halftoned image should resemble the local gray value. Here each dot is assumed to be localized in one point of the unit square. Within a region of constant grey value, the dots should hence be equally distributed. This can be achieved by minimizing a potential arising from a suitable combination of attraction and repulsion forces. Incorporating varying grey values then corresponds to locally adapting these forces. 

In this section, we review the mathematical foundations of this approach to halftoning.
For that,  we work with 
a continuous-domain model for grey-valued images given by a function $u: [0,1]^2 \mapsto [0,1] $. 
This includes pixeled images if we view them as piece-wise constant on small squares.



\subsection{The Attraction-Repulsion  Paradigm}

The idea to use an attraction-repulsion model for halftoning been first introduced in  \cite{Weickert2010} under the name electrostatic halftoning. 

\begin{figure}[ht!]
\begin{center}
  \subcaptionbox{ 
  \centering
  \label{fig121:a}}{\includegraphics[width=2.1in]{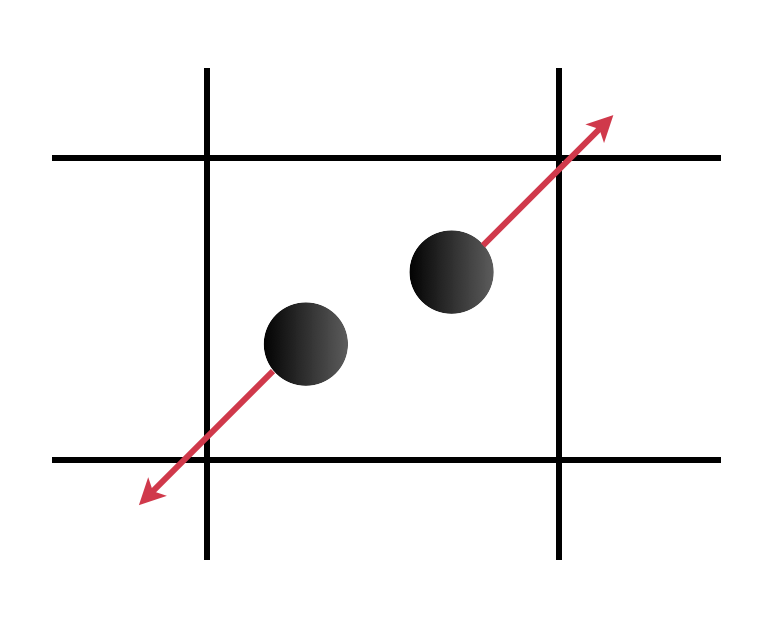}}
 \hspace{10mm}
  \subcaptionbox{  
  \label{fig121:b}}{\includegraphics[width=2.1in]{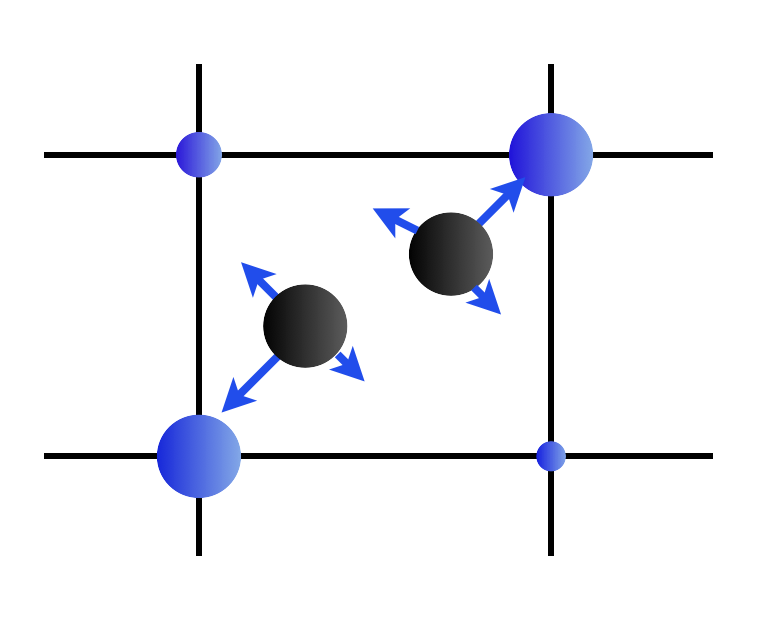}}%
 \caption{Illustration of attraction-repulsion model:
 (a) repulsion between particles,  (b) attraction by grid points.}
   \label{fig:Attraction:rep}
   
   \end{center}
\end{figure}

The idea is that electrostatic attraction forces scaling like the grey values of an image in combination with mutual repulsion (see Figure~\ref{fig:Attraction:rep} for an illustration) will drive a set of black points  $(\p_k)_{k=1}^m \in \R^2$ towards a distribution whose local averages resemble the image -- exactly as desired for halftoning.

 The first mathematical formulation of this approach is designed for pixeled images, so  we consider  a grid $\Gamma=\{1, \ldots, N\}\times \{1, \ldots, N\} $  and 
a grey-valued input image  $u: \Gamma \mapsto [0,1] $. Then, for a given number $m\in \N $ of black dots, we search for $m$ locations   $ (\p_k)_{k=1}^m= (p_{k,x}, p_{k,y})_{k=1}^m \in \R^2$ which
 best approximate the density described by $u$. By $\|\p_k\|:= \sqrt{p_{k,x}^2+p_{k,y}^2}$ we denote the Euclidean norm of the position of the $k$th black dot.

Initialized with a random set of locations $\big(\p^{(0)}_k\big)_{k=1}^m \in \R^2$, we view the dots as small charged particles in a 2-D particle system on a bounded domain evolving based on electrostatic laws. For constant grey values, the simulated evolution of the system leads to the maximization of relative distances between particles, resulting in a uniform distribution — a unicolor halftoned image.

For varying grey values, one extends the model by negative stationary charges at all grid points in $\Gamma$ proportional to the grey value of the corresponding pixel in the input image. These attractive forces guide the black dot locations toward the desired distribution representing the image.


Mathematically, the halftoning process can be described as follows.  Initializing, the positions of black dots $\big(\p^{(0)}_k\big)_{k=1}^m \in \R^2$ at random, one updates these positions in an iterative way by the update rule
\begin{align}\label{eq:Weickert-pixel-updates}
    \p^{(i+1)}_{k}&= \p^{(i)}_{k}+ \tau \Big( \sum_{\x \in \Gamma, \x\neq \p_{k} } \frac{1-u(\x)}{\|\x -\p_{k}\|} \ee_{k,\x} - \sum_{\x \in \Gamma,\p_{k} \neq \p_{m} } \frac{1}{\|\p_{m} -\p_{k}\|} \ee_{k,m} \Big)
\end{align}
where $\p^{(i)}_{k}$ denotes the location of black dots a at time level
$(i)$, and $\tau$ represents a small time step which is typically
chosen as $ \tau = 0.1 $. Moreover, $\ee_{k,m}$ and $\ee_{k,\x}$ are the unit vectors from the position $\p_k$ to the position $\p_m$ and from $\p_k$ to $\x$, defined respectively as
   $ \ee_{k,m} = \frac{\p_{m} -\p_{k}}{\|\p_{m} -\p_{k}\|}$ and $   \ee_{k,\x} = \frac{\x -\p_{k}}{\| \x -\p_{k}\|}. $

Here, one interprets  the first and the second sum in \eqref{eq:Weickert-pixel-updates} as two different forces
\begin{equation}
    \F^{(A)}_k= \sum_{\x \in \Gamma, \x\neq \p_{k} } \frac{1-u(\x)}{\|\x -\p_{k}\|} \ee_{k,\x}, \quad \quad \F^{(R)}_k=  \sum_{\x \in \Gamma,\p_{k} \neq \p_{m} } \frac{1}{\|\p_{m} -\p_{k}\|} \ee_{k,m}
\end{equation} 
The $\F^{(A)}_k$'s describe the attractive forces
originating from the discrete grid points, and the $\F^{(R)}_k$'s denote the repulsive forces between
particles. For more details about this interpretation, we refer
to Schmaltz et al. \cite{Weickert2010}. 

This comprehensive approach yields an aesthetically pleasing halftoning result, effectively synthesizing constant regions and edges in the image. Importantly, the overall system is designed to be electrically neutral, ensuring a near-optimal reproduction of the average grey value.  

However, there is a runtime issue associated with the initial electrostatic halftoning method. This issue is addressed in  \cite{Weickert2014} by introducing an improved version. The enhanced method is based on a novel nearest-neighbor identification scheme for continuous point distributions and uses Non-Uniform Fast Fourier Transform (NFFT) allowing for parallel processing with GPUs for faster computation. 
However, there were still no theoretical guarantees available for the algorithm's performance. 


Inspired by the original electrostatic halftoning method \cite{Weickert2010}, Teuber et al. \cite{Teuber2011} presented a related approach to halftoning, which focuses on global energy minimization. The proposed method involves minimizing a global energy functional, which is essentially the difference between two functions — where one is again related to dot attraction based on image grey values and the other, with a negative sign, associated with dot repulsion. More precisely, considering an image as a function $u: \Gamma \mapsto [0,1]$ on an integer grid $\Gamma=\{1, \ldots, n_{x}\}\times \{1, \ldots, n_{y}\} $ with its corresponding
weight distribution  $w:=1-u$,  the set of back point positions $\p:= (\p_k)_{k=1}^m= (p_{k,x}, p_{k,y})_{k=1}^m \in \R^2$ is to be determined as a  minimizer of the functional
\begin{equation}\label{eq:functional}
E(\p)=\sum_{k=1}^m\sum_{(i,j)\in \Gamma} w(i,j) \Big\| \p_k - {{i}\choose{j}}\Big\| - \lambda \sum_{k=1}^m\sum_{l=k+1}^m \|\p_k- \p_\ell\|, 
\end{equation}
where the parameter $\lambda$ is defined as some kind of averaged image weight~$
    \lambda= {\frac{1}{m} \sum_{(i,j)\in \Gamma} w(i,j)}.$
The functional \eqref{eq:functional},  consist of two functionals, the attraction functional $\F^{(A)}$ and the repulsion  functional $\F^{(R)}$ defined as
 \begin{equation*}
     \F^{(A)}(\p):= \sum_{k=1}^m\sum_{(i,j)\in \Gamma} w(i,j) \Big\| \p_k - {{i}\choose{j}}\Big\|   \quad \text{and}  \quad \F^{(R)}(\p):= - \sum_{k=1}^m\sum_{l=k+1}^m \|\p_k- \p_\ell\|
 \end{equation*}
respectively. Here, to minimizing $\F^{(A)}$, the black point  be attracted by the image
forces $w(i, j)$ at the points $(i, j) \in \Gamma $. On the other hand, there is a force of repulsion  $\F^{(R)}$ between
the particles modeled by the negative sign, which tries to maximize the distances between
the particles.



The key feature of the method proposed in \cite{Teuber2011} lies in formulating the functional as the difference of two convex functions (DC), allowing the authors to leverage existing results for computing minimizers of DC functionals. To understand the behavior of their functional, the authors initially explore the one-dimensional setting, demonstrating that the functional becomes convex under specific conditions. Furthermore, they show that minimizers can be determined analytically in this simplified setting. The authors establish that the minimizing points possess certain desired properties, such as being pairwise distinct, within the image frame, and positionable at grid coordinates. Some of these properties extend to the more complex two-dimensional case.


It is worth mentioning that the parameter $\lambda$ is not a usual regularization parameter but an equilibration parameter between the “opposite” functionals. Since $\frac{1}{\lambda} E(\p)$ has the same minimizer as $E(\p)$, the authors  propose setting $\lambda := 1$ and using  $\frac{1}{\lambda}w(i, j)$ instead of $w(i, j)$ in the functional. Therefore, one can restrict one's attention mainly to $\lambda := 1$  and ${m =\sum_{(i,j)\in \Gamma}
 w(i, j)}$. 
 This paper is also the first that discusses the discretization issue of digital halftoning: While in many applications one is interested in digital halftoning with dots lying on a pixel grid, electrostatic halftoning is an analog halftoning scheme and hence, in general, yields continuous coordinates.
 As a remedy, the authors propose to round the coordinates to the nearest integer locations. The effects of this rounding on the representation quality, however, are not investigated mathematically. For the total number of black dots $m$, they propose the number $m:= round(\sum_{(i,j)\in \Gamma} w(i, j))$ as it preserves the mean gray value of the image.
As the authors argue, their novel approach to halftoning also generalizes to the stippling problem, where one does not use black dots, but line segments as building blocks.


\subsection{From Attraction-Repulsion Models to Quadrature Rules and Discrepancies in RKHSs}

To provide a corresponding theory for the two-dimensional case, the paper \cite{graf2012quadrature} relates the attraction-repulsion model for halftoning to quadrature errors in a reproducing kernel Hilbert space (RKHS).
The authors consider 
a more general version of the functional \eqref{eq:functional} as given by
\begin{equation}\label{eq:functional:cont}
E(\p)=\sum_{k=1}^M\int_{\X} w(x) \phi(\| \p_k - x\|_2) \dx x - \lambda \sum_{k=1}^M\sum_{l=k+1}^m \phi(\|\p_k- \p_\ell\|_2), 
\end{equation}
with some function $\phi:[0,\infty] \to \R$. For example, $\phi(x):= x$ leads exactly to the continuous analog of  \eqref{eq:functional}. 


Before summarizing their investigations, we will briefly recall the concepts of an RKHS and the quadrature error. 

For $\X\in \{\R^d, [0,1]^d, \S^1, \T^2, \S^2\}$,  
a real reproducing kernel Hilbert space (RKHS), $H_K$,  is a Hilbert space having a reproducing kernel, a function ${K: \X \times \X \to \R}$ which fulfills 
\begin{equation}
    K_x:=K(\cdot, x) \in H_K, \quad \text{and} \quad  f(x)=\langle f, K(\cdot, x) \rangle \quad \forall x \in \X, \; \forall  f \in H_K. 
\end{equation}
For more information on RKHSs, we refer to \cite{aurenhammer1998minkowski}.

In the following, the authors assume that the image is modeled by a continuous function $w:\X \to \R$; when $\X=\R^d$, one additionally assumes that $w$ has compact support. Furthermore, for the kernel $K$ it holds that 
\begin{equation}
    h_w=\int_{\X}w(y) K(x,y)\dx y \in H_K
\end{equation}
i.e. $\|h_w\|^2_{H_K}=\int_{\X}\int_{\X}w(x)w(y) K(x,y)\dx x\dx y< \infty.$

The key observation of \cite{graf2012quadrature} is that the problem of minimizing the energy $E_K(\p)$ over potential black dot locations $\p := (\p_1, \ldots, \p_M) \in \X^M$, is closely related to the problem of approximating integrals of the form
\begin{equation}
    I_{w}(f)=\int_{\X} f(x) w(x) \dx x\quad \text{for} \quad f\in H_K
\end{equation}
by a quadrature rule
\begin{equation}
    Q(f, \p)=\lambda \sum_{i=1}^M f(\p_i) \quad \lambda = \frac{1}{M} \int_{\X}w(x) \dx x.
\end{equation}

Quantitatively, this dependence will be expressed via the worst-case quadrature error as given by
\begin{equation}\label{eq:quadrature}
err_k(\p):= \sup_{f\in H_K, \; \|f\|_{H_K}\le 1} |I_w(f)- Q(f, \p)|= \|I_w- Q(\cdot, \p)\|,
\end{equation}
where the later norm is the operator norm of the linear functionals on $H_K$. More precisely, the attraction-repulsion functional \eqref{eq:functional:cont}  leads to the same optimal point distributions as the quadrature error functional for a certain
RKHS of functions on $\R^2$ with the Euclidean distance kernel.
\begin{thm}\cite{graf2012quadrature}
    Let K be a positive semi-definite function and $H_K$ the associated RKHS. Then
the relation
\begin{equation}
    err_k(\p)^2=2 \lambda E_K(\p) + \|h_w\|^2_{H_K}
\end{equation}
holds true, where
\begin{equation}\label{eq:quad:error:fun}
    E_K(\p):=\frac{\lambda}{2} \sum_{i,j=1}^M K(\p_i, \p_j) - \sum_{i=1}^M \int_{\X} w(x) K(\p_i, x)\dx x.
\end{equation}
\end{thm}

Taking in \eqref{eq:quad:error:fun} a radial kernel  $K : \X\times \X \to \R$  with $K(x,y)=\phi(\|x-y\|_2)$ with general functions $\phi:[0,\infty] \to \R$, we arrive exactly at the negative  of functional \eqref{eq:functional:cont}.

Computation of minimizers of \eqref{eq:quad:error:fun} is, in general, a challenging problem itself. Therefore, the authors in \cite{graf2012quadrature} also developed algorithms for the efficient computation of minimizers $\Hat{\p}$ of functionals $E_K$ for given functions $w$ on $ \S^1$, $\T^2$ and $\S^2$ using a nonlinear conjugate gradient method and the fast Fourier transforms at nonequispaced nodes on the torus and the sphere. 

The connection to the quadrature rule also sheds some light on why the attraction-repulsion model aligns with our goal of providing dot configurations whose low-frequency content agrees with that of the image to be represented (recall that this objective is motivated by modeling human vision via a low-pass filter).  To see that, let $\X\in \{\S^1, \T^2, \S^2\}$ and ${\phi_l: l\in \N}$ be an orthonormal basis of $L_2(\X)$. Then any  real-valued function $w\in L_2(\X)$ representing an image can be written in the form
\begin{equation}
w(x)=\sum_{l=1}^\infty \hat{w}_l\phi_l(x), \quad \quad \hat{w}_l=\langle w, \phi_l \rangle_{L_2}=\int_{\X} w(x) \overline{\phi_l(x)} \dx x.
\end{equation}

Working in the quadrature rule scenario with bandlimited functions $w$, i.e. $\hat{w}_l=0$ for $l\le d_N$ for a bandwidth $N\in \N$, and applying bandlimited kernels of the form
$$
K_N(x,y)=\sum_{i=1}^{d_N} \lambda_l \phi_l(x)\overline{\phi_l(y)} 
$$
with $ \lambda_l>0$ allows to represent the quadrature error $err_k(\p)^2$ as x
\begin{equation}
    err_k(\p)^2= \sum_{l=1}^{d_N} \lambda_l \left| \lambda  \sum_{i=1}^M \overline{\phi_l(\p_i)} -  \hat{w}_l\right|^2. 
\end{equation}
This representation of the quadrature error shows that the image $w$ and its black dot representative are close in the low-pass domain exactly when $err_k$ is small, so the attraction-repulsion model is in line with the main idea of image halftoning. 

Moreover, in the same work \cite{graf2012quadrature}, the authors show that the quadrature errors considered above are closely related to so-called discrepancies, which adds another interesting point of view on the halftoning process.    

To reflect this point of view, we consider the following setup. Let $\X \in \{\R^d, [0,1]^d, \S^1, \T^2, \S^2\}$ and $D:=\X\times [0,R]$ for some fixed $R\in \R_+$ and let $\B(c,r):=\{x \in \X: d_{\X}(c,x)\le r\} $ be the ball centered at $c \in X$ with
radius $0 \le r\le  R.$ Define a positive semi-definite kernel $K_\B$ as 
\begin{equation}\label{eq:discrepancy-kernel}
    K_{\B}(x,y):= \int_{O}^R\int_{\X} \Ind_{\B(c,r)}(x)\Ind_{\B(c,r)}(y)\dx c \dx r, 
\end{equation}
where $\Ind_{\B(c,r)}$ denotes the characteristic function of $\B(c,r)$.
The $L_2$-discrepancy between the set of  black dot locations $\p := (\p_1, \ldots, \p_M) \in \X^M$ and the  ground truth image $\omega$ with respect to this kernel is defined as
\begin{equation}\label{eq:discrepancy-1st-def}
    disc_2^{\B}(\p):= \left( \int _{D} \left(\int_{\X} \omega(x)\Ind_{\B(t)}(x)\dx x - \lambda \sum_{i=1}^M \Ind_{\B(t)}(\p_i)\right)^2\right)^{\frac{1}{2}},
\end{equation}
where we denote $t:=(c,r)\in D$, $\dx t:=\dx c \dx r$.
Then one obtains the following connection:
\begin{thm}
    Let $K_B$ be defined by \eqref{eq:discrepancy-kernel} and let $H_{K_B}$ be the associated RKHS of functions
on $\X$. Then values of the quadrature rule $err_{K_{\B}}$ given by \eqref{eq:quadrature} and $disc_2^{\B}$ determined above coincide 
\begin{equation*}
err_{K_{\B}}(\p)= disc_2^{\B}(\p).
\end{equation*}
\end{thm}

As we see in the following subsection, discrepancies are also closely related to image low-pass filtering.

\subsection{Attraction-Repulsion and Discrepancies Models for Images as Probability Measures}

Building on the main motivation of electrostatic halftoning to place the dots so that their distribution approximates the gray values, the paper \cite{Fornasier2016} studies the limiting case of a number of particles growing to infinity.  The first challenge is to precisely formulate a suitable concept of convergence. Namely, at any approximation level, there will still be a finite number of dots, so many concepts of convergence do not apply. In \cite{Fornasier2016}, this issue is addressed by interpreting both the image and its halftoned representation as measures -- the former a measure $\mu$ with density $w$, the latter a sum $\mu_N$ of $N$ Diracs -- and then aim to establish the so-called narrow convergence of the halftoned representations to the image.
Here $\mu_N$ converges narrowly to $\mu$ if one has
\begin{equation}
    \lim_{N\to \infty}\left| \int_{\X}g(x)\dx \mu_N(x) - \int_{\X}g(x)\dx \mu(x)\right|=0,    
\end{equation}
 for all functions $ g\in C_b(\X)$, where $C_b(\X)$ is the space of bounded continuous functions on $\X$. Narrow convergence is in line with the intuition that one aims for approximation after applying a filter -- in this case $g$. For more details on the narrow topology in spaces of probability measures $\Pp(\X)$, we refer the readers to  \cite{ambrosio2005gradient}.       



More precisely, a black-and-white image on a domain ${\X\subset \R^d}$ is modeled by a measure in the space of probability measures $\Pp(\X)$  whose density is proportional to the local gray values. 
In the halftoned image, the black dots are modeled as Dirac delta measures,  which are then averaged to obtain 
$$\mu_N=\frac{1}{N}\sum_{i=1}^N\delta_{\p_i} \in \Pp(\X), $$
where $\p = (\p_i )_{i=1}^N\subset \X$ denotes the set of dot locations. 

The dot locations are again obtained by minimizing an attraction-repulsion energy. 
Refining the previous approaches, the authors allow the use of different kernels for attraction and repulsion in functional \eqref{eq:quad:error:fun}. One considers kernels of the form ${K_{(A)}(x,y):=\psi_{(A)}(x-y)}$ and $K_{(R)}(x,y):=\psi_{(R)}(x-y)$, respectively,   with radially symmetric
functions $\psi_{(A)}, \psi_{(R)}\colon \R^d \to \R_+$.  The resulting energy then takes the form
\begin{align}\label{eq:attrepmassimo}
    E^{(A,R)}_N[\p]&= \int_{\X\times \X} \psi_{(A)}(x-y) \dx \omega(x)\dx \mu_N(y) - \frac{1}{2} \int_{\X\times \X} \psi_{(R)}(x-y) \dx \mu_N(x)\dx \mu_N(y)\\
    &= \frac{1}{N}\sum_{i=1}^N  \int_{\X}\psi_{(A)}(\p_i-y) \dx \omega(x) - \frac{1}{2} \frac{1}{N^2}\sum_{i,j=1}^N  \psi_{(R)}(\p_i-\p_j). 
\end{align}
Indeed, the main result of \cite{Fornasier2016} establishes that the minimizer of this functional has the desired narrow convergence properties provided the probability measures under consideration have bounded second moment and $\psi_{(A)}$ as well as $\psi_{(R)}$ exhibit at most quadratic growth. Examples of admissible radially symmetric functions hence include $\psi_{(A)}(\cdot)=\psi_{(R)}(\cdot)= \|\cdot\|^q$ for $1\le q\le 2$.


The proof of this result is based on a connection between the narrow convergence and the so-called $\Gamma$-convergence \cite{dal1993progress}, a key convergence concept in functional analysis, in combination with a suitable Fourier representation of the attraction-repulsion model \eqref{eq:attrepmassimo}.




    

As an alternative measure for the approximation quality  of images modeled as probability measures, the authors of
 \cite{ehler2021curve} consider a version of the discrepancy (cf.~\eqref{eq:discrepancy-1st-def} above). To precisely state this for probability measures, we return to the RKHS setting. Consider a continuous, symmetric, positive definite function $K: \X \times \X\to \R$. By Mercer's theorem \cite{cucker2002mathematical}
there exists an orthonormal basis $\phi_k, \, k\in \N$  of $L^2(\X, \sigma_{X})$ and non-negative coefficients
$(\alpha_k)_{k\in N}\in \ell_1 $ such that $K$ has the Fourier expansion
\begin{equation}
    K(x,y)=\sum_{k=0}^\infty\alpha_k \phi_k(x)\overline{\phi_k(y)}
\end{equation}
with absolute and uniform convergence of the right-hand side. Moreover, every function $f\in L^2(\X, \sigma_{X})$ has a Fourier
expansion
$  f=\sum_{k=0}^\infty \hat{f}_k \phi_k, \quad  {f}_k:=\int_{\X}f\overline{\phi_k}\dx \sigma_{\X}. 
$
The reproducing kernel Hilbert space associated with the kernel $K$ is denoted as 
\begin{equation}
    H_K(\X)=\{f \in L^2(\X, \sigma_{X})\colon \, \sum_{k=0}^\infty \alpha_k^{-1} |\hat{f}_k|^2<\infty \}. 
\end{equation}

With these notions, we can now define the discrepancy:
For two probability measure $\nu, \mu \in \Pp(\X)$, the discrepancy $disc(\nu, \mu)$ is defined as the dual norm on $H_K(\X)$ of the linear operator
$T : H_K(\X) \to  \C$  with $\phi \mapsto \int_{\X} \phi \dx ( \nu- \mu)$, namely
\begin{equation*}
    disc(\nu, \mu)= \max_{\|\phi\|_{H_K(\X)}\le 1} \left| \int_{\X} \phi \dx ( \nu- \mu) \right|. 
\end{equation*}

The authors show that $disc(\nu, \mu)^2$ can be equivalently represented in terms of the Fourier coefficients of  $\nu, \mu \in \Pp(\X)$ as 
\begin{equation*}
    disc(\nu, \mu)^2=\sum_{k=0}^\infty \alpha_k |\hat{\mu}_k-\hat{\nu}_k|^2, 
\end{equation*}
where the Fourier coefficients of $\nu, \mu \in \Pp(\X)$ are defined as $\hat{\mu}_k= \int_{\X}f\overline{\phi_k}\dx \mu$ and $\hat{\nu}_k=\int_{\X}f\overline{\phi_k}\dx \nu$ respectively, and $\alpha_k$ are the coefficients in the kernel expansion.
As the $\alpha_k$'s decay, this puts a higher weight on the low-pass part. This again resembles the main idea of image halftoning stated in the beginning, namely that the true image and the halftoned image should look similar in the low-pass band.


With this notion of similarity, the authors of \cite{ehler2021curve} study a somewhat more general setup: They examine how well a probability measure $\omega\in \Pp(\X)$ can be approximated by a probability measure supported on Lipschitz curves, which correspond to push-forward measures of the Lebesgue measure on the unit interval onto such curves. Note that for many spaces $\X$, this class of push-forward measures also includes the averaged Dirac measures discussed above, so the scenario also covers halftoning.

To precisely formulate this, denote by $C([a; b];\X)$ denote the set of closed, continuous curves 
 $\gamma\colon[a; b] \to 
\X$ and recall that the
length of a curve $\gamma \in C([a; b];\X)$ is given by
\begin{equation}\label{eq:length-cure}
\ell(\gamma):=\sup_{{a\le t_0\le \ldots\le t_n\le b, n\in \N}} \sum_{i=1}^N dist_{\X}(\gamma(t_i),\gamma(t_{i+1})).
\end{equation}
With this notation, the space of measures supported on Lipschitz
continuous curves of length $L$ in $\X$ is given by
\begin{equation*}
    \Pp_L^{curv}(\X):= \{\nu \in \Pp(\X): \exists \gamma \in C([a; b];\X), \, \supp(\nu)\subset \gamma([a,b]), \, \ell(\gamma)\le L\}. 
   \end{equation*}
The main result of \cite{ehler2021curve} establishes that measures in $\Pp_L^{curv}(\X)$ are able to approximate general probability measure at a rate decaying with the length of the supporting curves: 
\begin{thm}
    For any $\mu\in \Pp(\X)$, it holds with a constant $C>0$ depending on $\X$ and kernel $K$ that 
    \begin{equation*}
        \min_{\nu \in \Pp_L^{curv}(\X)} disc(\nu, \mu) \le C \cdot L^{-\frac{d}{2d-2}}. 
    \end{equation*}
\end{thm}

 The numerical minimization of the discrepancy between a
given probability measure and the set of push-forward measures of Lebesgue
measures on the unit interval by Lipschitz curves \cite{ehler2021curve} is handled here by an algorithm based on a conjugate gradient method on manifolds, the performance of which is showcased for  measures on the 2- and 3-dimensional torus, the 2-sphere, the
rotation group on $\R^3$ and the Grassmannian of all 2-dimensional linear subspaces of $\R^4$. 
The numerical minimization requires the discretization of curves as potential measure supports by a set of $(\p_k)_{k=1}^m$, which relates back to placing black dots in a setting similar to halftoning.   




\section{Halftoning via Error Diffusion Schemes}\label{sec:error-diff-halftoning}

A drawback of attraction-repulsion or discrepancy-based methods is that the positions of black points $(\p_k)_{k=1}^m$ obtained at the end of the update rule \eqref{eq:Weickert-pixel-updates} or from minimizing functional \eqref{eq:functional}, are in most cases not positioned on the pixel grid $\Gamma$. Therefore after the minimization procedure, one has to round the points to grid points or enforce the constraint in other ways. To our knowledge, the effects of these discretization steps are not understood, and it is also not clear what is the best strategy. This may be a reason why in many applications, conceptually very different methods are applied, namely error diffusion schemes.

The main idea remains the same: one aims to arrange black pixels to mimic the distribution of gray values. However, the strategy applied is local rather than global: the idea is to distribute the halftoning error that occurs during the conversion process to neighboring pixels so that the errors average out. This is done by tracking an accumulated error parameter that is updated via a linear recurrence relation combined with rounding when passing spatially over the image. This approach diffuses the error over the image, 
yielding a visually pleasing halftoned representation. This approach has been demonstrated to reduce noticeable artifacts compared to  previous approaches such as ordered dithering \cite{limb1969design, lippel1971ordered}
 and random dithering \cite{purgathofer1994forced} techniques. A popular example of an error diffusion technique is the Floyd–Steinberg \cite{SteinbergFloyd} scheme, notable extensions can be found in \cite{Knuth1987, Jarvis1976, Jarvis1976-2, Shiau1996, billotet1983error, levien1992output, eschbach1997error, kite2000modeling}.
However, the  parameters for the recurrence relations in these works were mainly determined empirically, and only very recently, first steps towards a mathematical understanding were taken \cite{Krahmer2023-siam, Krahmer2023} building on a connection
to $\Sigma\Delta$  quantization of band-limited functions as first observed in \cite{Knox1992, Kite1997}. 
Namely, the black-and-white pixels in the halftoned image can be interpreted as bits of a quantized representation, and the quality measure based on comparing low-pass components is in direct analogy to the Shannon sampling theorem in signal processing.

In this section, we will first 
review different halftoning approaches as they have been proposed in the imaging literature without much theoretical underpinning. We will then discuss their connection to Sigma-Delta quantization.
For that, we first introduce the basics of Sigma-Delta quantization for bandlimited signals. After that, we will explain the connection and discuss how it can be exploited for the design of error diffusion schemes.

\subsection{Classical Error Diffusion Schemes for Image Halftoning}
In this section, we work with fully discrete images given by a configuration of pixels $ \p: =\lbc \p_{n_1,n_2} \rbc_{n_1, n_2=0}^{N}$ with $\p_{n_1,n_2}\in [-1,1]$. Note that for reasons of consistency with the following parts we use the unusual rescaling to a centered interval (other than many papers, but of course, the representations are entirely equivalent).
Consequently, we are looking for a halftoned representation  $\q:=\lbc \q_{n_1,n_2} \rbc_{n_1,n_2=0}^{N}$ that satisfies $ \q_{n_1,n_2}\in \{-1,1\}$.






\begin{figure}
\begin{center}
	\includegraphics[width=0.45\textwidth]{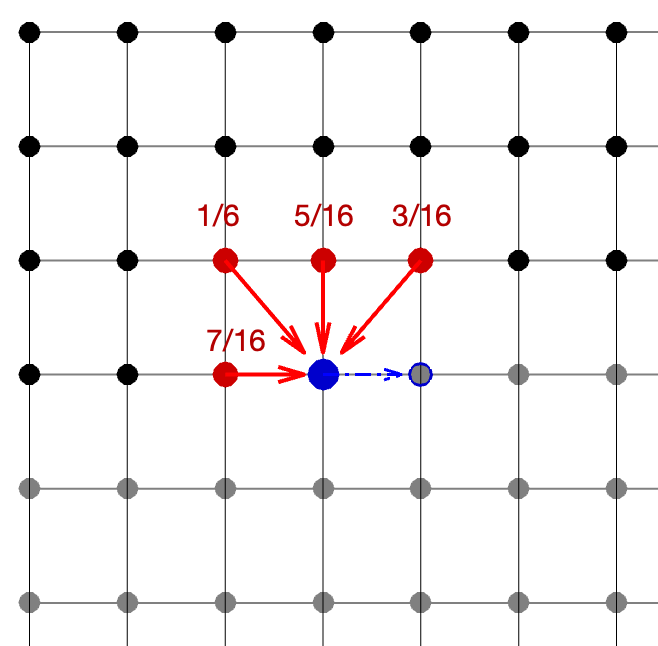} 
	\caption{\small  The elements of $\vv$ used (in red) at current quantization step $(n_1,n_2)$ (in blue) to define $\vv_{n_1,n_2}$ for  Floyd–Steinberg $\SD$ halftoning scheme. Black points denote already half-toned elements and the next step is marked by the grey disk with blue bounds. }
		\label{fig:subim1}
		\end{center}
\end{figure}
The idea of error diffusion is to compute $\q$ via a recurrence relation based on keeping track of an accumulated error quantity.
The first such approach was the celebrated Floyd–Steinberg halftoning algorithm \cite{SteinbergFloyd} introduced in 1976, in which the halftoned image $\q$ and the accumulated error sequence $\vv$ are obtained via the  following iterative scheme
\begin{align}\label{SF-scheme-1}
    \q_{m,n}&=\sign\Big( \tfrac{5}{16} \vv_{m-1,n}\!+\!\tfrac{7}{16}\vv_{m,n-1}\!+\!\tfrac{1}{16}\vv_{m-1,n-1}\!+\!\tfrac{3}{16}\vv_{m-1,n+1}\!+\!\p_{m,n}\Big) \\
     \vv_{m,n}&=  \tfrac{5}{16} \vv_{m-1,n}\!+\!\tfrac{7}{16}\vv_{m,n-1}\!+\!\tfrac{1}{16}\vv_{m-1,n-1}\!+\!\tfrac{3}{16}\vv_{m-1,n+1}\!+\!\p_{m,n}-\q_{m,n},\label{SF-scheme-2} 
\end{align}
 where the sign function is given as \begin{equation}
    \sign(x):= \lbc \begin{matrix} 1,& x>0, \\
    -1, & x\le 0.\end{matrix} \right.
\end{equation}
The halftoned values at which neighboring pixels are taken into account in the recurrence relation and with which coefficients are by no means obvious and multiple alternative error diffusions with different recurrence relations have been proposed
 \cite{Knuth1987, Jarvis1976, Jarvis1976-2, Shiau1996, billotet1983error, levien1992output, eschbach1997error, kite2000modeling}. 

Most of these approaches including the Floyd-Steinberg algorithm pass through the image line by line, so they can only take into account the halftoned values both in the current line left of the pixel to be computed and in the entire lines above in the recurrence relation, see Figure~\ref{fig:subim1} for a visualization for the Floyd-Steinberg scheme. A notable exception is \cite{Shiau1996}, where also other ways to pass through the image, e.g.~on zigzag paths, are explored.


A challenge in comparing these approaches and developing them further is how to quantify the quality of the obtained representation. Most papers on the subject have used individual visual inspection as a measure, for example comparing the representation of fine details. A more quantitative approach has been proposed in \cite{Pang2008}, where the authors propose to use similarity indices, but still this measure does not shed any light on why certain
halftoned images are better than others and how can we measure the quality of a halftoned image. The first steps towards an answer \cite{Krahmer2023-siam, Krahmer2023}  are based on the connection to Sigma-Delta quantization; to present these results, we first recall the basics of the $\SD$ theory.

\subsection{Basics of Sigma-Delta Quantization}

The goal of $\SD$ quantization is to represent a bandlimited signal by a sequence of bits whose low-pass content agrees with the signal.
Here a bandlimited signal is one that is compactly supported in the frequency domain.
Without loss of generality, we assume that this compact support is the centered unit square ${S_1:=[-\frac{1}{2},\frac{1}{2} ]\!\times\![-\frac{1}{2},\frac{1}{2}]\subset\R^2}$.

More precisely, the class of $S_{1}$-bandlimited functions $\B_{S_{1}}$, is the set of real-valued continuous functions in $L^\infty(\R^2)$ 
whose Fourier transform  given by $\mathcal{F} w( \bm \xi)= \int_{\R^2} w(\bm x)\e^{-2\pi\im \bm \xi \cdot \bm x} \dx \bm x $ (extended to the space of tempered distributions in the natural way) exists and  
vanishes outside of the region $S_{1}$. 
 Shannon's sampling theorem \cite{Petersen} states that such functions $w\!\in\!\B_{S_{1}}$  can be reconstructed  from its samples on the lattice $\!\frac{1}{\lambda}\,\Z^2$ with an oversampling rate $\lambda\geq 1$,  via the sampling formula 
\begin{equation}\label{sampling2D}
     w (\x)=\frac{1}{\lambda^2}\sum\limits_{\bm n\in \Z^2} w\Big(\frac{\bm n}{\lambda}\Big)\Phi\Big( \x- \frac{\bm n}{\lambda}\Big), 
\end{equation}
where $\Phi$ is a sinc-kernel.
When $\lambda>1$, one can also choose $\Phi$ as a Schwartz function with the low-pass property 
  $ {\mathcal{F}\Phi(\bm \xi)=\left\{\begin{matrix}
    1,& \bm \xi \in S_{\lambda},\\
    0,&\bm \xi \notin S_{\lambda}.\end{matrix}\right.}$,
    thus having considerably better decay properties.
To restrict to a fixed amplitude range, we consider the function class  $ \B^b:= \lbc w\in \B_{S_1}\colon \; \nofty{w}\le b \rbc$. 

The goal is now to replaces the samples $w\Big(\frac{\bm n}{\lambda}\Big)$ in the sampling formula by values $q_n$ from a finite alphabet such that the quantized representation
\begin{equation}
  w_{q}(\x)= \frac{1}{\lambda^2}\sum\limits_{\bm n\in \Z^2} q_{\bm n}\Phi\Big( \x- \frac{\bm n}{\lambda}\Big), \quad \quad  \x \in \R^2_{+}. 
\end{equation}
is close to $w$ in a suitable metric. In this article, we focus on the error metric $\|e\|_{L^{\infty}(\R^2_+)}$, where $e$ is the error signal (or error function)  given by $e(\x):= w(\x)- w_{q}(\x),  \; \x \in \R^2_{+}$. 
Furthermore, motivated by the halftoning application for grey-valued images, which only admits black and white pixels, we focus on the one-bit alphabet $\mathcal{A}=\{-1,1\}$.

As proposed in  \cite{Gunturk2003}, one can
decompose the error signal  into two terms $e=e_{w}+e_{q}$, where
$ e_{w}(\x):= w(\x)- w_{\lam}(\x), $ $  e_q(\x):= w_{\lam}(\x)-w_q(\x) $ and 
\begin{equation}
  w_{\lam}(\x)= \frac{1}{\lam^2}\sum\limits_{\bm n\in \N^2} w\Big(\frac{\bm n}{\lam}\Big)\Phi\Big( \x- \frac{\bm n}{\lam}\Big), \quad \quad  \x \in \R^2_{+}.
\end{equation}
One can then argue that the term $e_q$ is dominant away from the boundary, so one needs to focus only on $e_q$ as a halftoning error. 

In one dimension, the problem of minimizing $e_q$ has been extensively studied in the literature. A class of  particular relevance are so-called $\SD$ quantizers used  to quantize the samples $p_{n}:=w(\frac{n}{\lam})$ 
of univariate bounded and bandlimited function  $w\in \B^1$, where $\lambda>1$ denotes the oversampling rate. A $1$-bit $\SD$ quantizer then outputs a sequence $q=\{q_n\}_{n\in \N}$ with  $q_{n}\!\in\!\{-1,1\}$ obtained by solving the difference equation
\begin{align}
  v_{n}&= (h*v)_{n}+ p_n-q_n \label{1D-SD-def-1}\\ 
  q_n&= \sign\big((h*v)_{n}+y_{n}\big), \label{1D-SD-def-2}
\end{align}
where the feedback filter  $h\in \ell^1$ satisfies $h_n=0$ if $n\le 0$,  the state variable $v_n$ is set to zero for $n<0$ \cite{Gunturk2003},
and the sign function is given as  
\begin{equation}
    \sign(x):= \lbc \begin{matrix} 1,& x>0, \\
    -1, & x\le 0.\end{matrix} \right.
\end{equation}
 
For a positive integer $r$ and a sequence $g\in \ell^1$ with $g_n=0$ for $n<0$, the $\SD$ {\it quantizer} \eqref{1D-SD-def-1}-\eqref{1D-SD-def-2} is called {\it of order $r$} as soon as the filter $h$ satisfies the identity \begin{equation}\label{fil-def}
    \delta^0-h=\Delta^r g, 
\end{equation} 
where $\delta^a$ denotes the Kronecker delta sequence situated at the integer $a$.
As shown in \cite{Gunturk2003},   via the change of variables $u=g*v$, \eqref{1D-SD-def-1} can then be expressed in the more classical form $
   (\Delta^ru)_n= p_n-q_n$. 

  In the one-dimensional case, the quantization process has been well understood \cite{Daubechies2003, Gunturk2003, deift2011optimal}, and it has been shown that the error $e_q$ for an $r$th-order scheme diminishes with the oversampling rate $\lam$ at a rate of   $\mathcal{O}({\lam}^{-r})$. By appropriately combining schemes of different order, one can even achieve exponential error decay in $\lambda$ \cite{Gunturk2003, deift2011optimal}, see \cite{Krahmer2012} for corresponding lower bounds.
  It should be noted that these lower bounds strongly increase when the signal amplitude approaches $1$ -- in fact, all schemes proposed beyond first order require an amplitude that is strictly bounded away from $\pm 1$. In the image representation context, this would correspond to a low-contrast image that neither has fully black nor fully white regions -- arguably not realistic for real-world grey-valued images.

\subsection{Error Diffusion Schemes as Weighted $\SD$ Quantization Schemes}

% ... 


The connection between $\SD$ quantization and halftoning is based on the fact that the Shannon sampling formula is nothing but a low-pass filter applied to the samples. So when the low-pass operation that approximately describes the smoothing effect in human vision is taking effect on a halftoned image, this is directly analogous to reconstructing a bandlimited signal from quantized samples. 

Nevertheless, the mathematical understanding of $\SD$ quantization only partly applies to halftoning, as the latter is intrinsically two-dimensional, but $\SD$ has mainly been studied in one dimension.

That said, the formulation of the problem can easily be extended to 2D:
For that, we begin with the definition of the finite-difference operator in the bivariate case. Recall that  the (backward) finite difference $\Delta$ operator maps a 1D-sequence $v=\{v_n\}_{n \in \Z}$ to the sequence  $\Delta v= \{(\Delta v)_{ n}\}_{n \in \Z}$ with $(\Delta v)_{ n}=v_n-v_{n-1}$. Consequently, the $r$th order finite difference operator $\Delta^r$ is defined via 
$  (\Delta^r v)_{ n}=\sum_{j=0}^r (-1)^{j}{\tiny {{r}\choose{j}}}   v_{n-j} 
$
In the two-dimensional case, finite difference operators can be defined in arbitrary directions. Namely, on the lattice  $\L\in \R^2$, for a direction $\bb d= (d_1, d_2)\in \L$ and a sequence 
${v=\{v_{\bm n}\}_{\bm n\in \L}}$,  
 the sequence $\Delta^r_{\bb d}v= \{(\Delta^r_{\bb d}v)_{\bm n}\}_{\bm n\in \Z^2}$ of its  directional finite (backward) difference of order $r$ is defined as 
\begin{equation}\label{DD-def-of-SD}
    (\Delta^r_{\bb d}v)_{\bm n}=\sum_{j=0}^r (-1)^{j}{\tiny {{r}\choose{j}}}   v_{\bm n-j\bb d}. 
\end{equation}

Analogously, one defines  directional convolution for 
 a given one-dimensional filter $h=\{h_j\}_{j\in \Z}$ supported on the first $L$ elements and a two-index sequence $v=\{v_{\bm n}\}_{\bm n\in \Z^2}$, we denote by $h*_{\bb  d}v$  the  convolution of $v$ and $h$ in the direction $\bb d$ and define it as
\begin{equation}
    (h*_{\bb d }v)_{\bm n}= \sum\limits_{j=1}^Lh_j v_{\bm n -j \bb d}
\end{equation}
With this notion, the directional finite difference can be expressed as $(\Delta^r_{\bb d}v)=\Delta^r*_{\bb d } v $ where $\Delta$ denotes the sequence given by  $\Delta_0=1$, ${\Delta_1=-1}$, ${\Delta_k=0}$, for all ${k\in \Z\setminus{\{0,1\}}}$, and $\Delta^r:=\Delta*\ldots*\Delta$.

With these definitions, one observes that the Floyd-Steinberg error diffusion method \eqref{SF-scheme-1} is nothing but a weighted average of first-order one-dimensional $\SD$ schemes applied in the four directions ${\bb d_{1,0}\!=\!(1,0)}$, $\bb d_{1,1}\!=\!(1,1),$ $ \bb d_{0,1}\!=\!(0,1),  {\bb d_{-1,1}\!=\!(-1,1)}$ represented  now as 
\begin{align*}
    \tfrac{7}{16}(\Delta_{\bb d_{0,1}}\vv)_{m,n}&\!+\!\tfrac{1}{16}(\Delta_{\bb d_{1,1}}\vv)_{m,n}\!+\!\tfrac{5}{16}(\Delta_{\bb d_{1,0}}\vv)_{m,n}\!+\!\tfrac{3}{16}(\Delta_{\bb d_{-1,1}}\vv)_{m,n} =  \p_{m,n}\!-\!\q_{m,n}\\ 
q_{m,n}&=\sign\Big( \!\vv_{m,n}\!-\!\tfrac{7}{16}(\Delta_{\bb d_{0,1}}\vv)_{m,n}\!-\!\tfrac{1}{16}(\Delta_{\bb d_{1,1}}\vv)_{m,n}\!-\!\tfrac{5}{16}(\Delta_{\bb d_{1,0}}\vv)_{m,n}\\
&\quad \quad \quad \quad  \quad  \quad \quad  \quad  \quad \quad  \quad  \quad \!-\! \tfrac{3}{16}(\Delta_{\bb d_{1,-1}}\vv)_{m,n}\!+\!\p_{n,m}\Big)
\end{align*}

\begin{figure}
    \centering
    \includegraphics[width=0.45\textwidth]{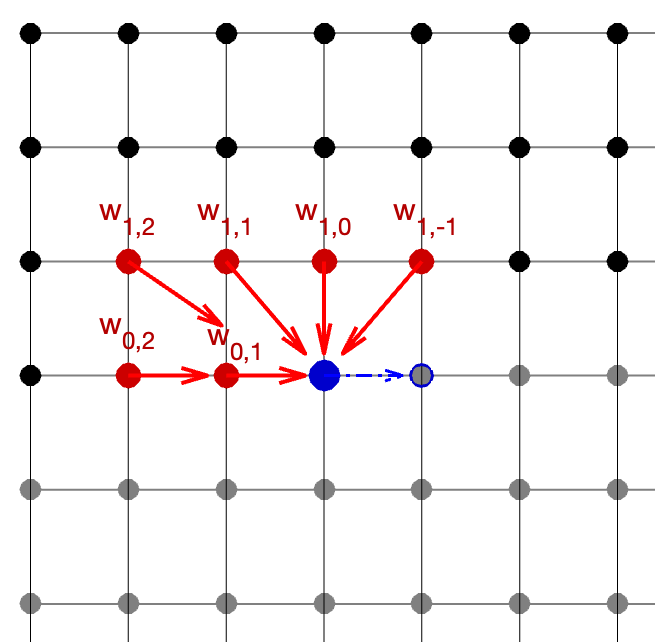}
    \caption{ Weighted elements of $\vv$ (in red) used at current quantization step $\bm n$  to define $v_{\bm n}$ (in blue) for the  weighted  $\SD$ quantization scheme with $\W\in \R^{4\times 2}$. Black points denote already quantized elements and the next step is marked by the grey disk with blue bounds.}
    \label{fig:my_label}
\end{figure}

In fact all the proposed in the literature error-diffusion methods \cite{Knuth1987, Jarvis1976, Jarvis1976-2, Shiau1996, billotet1983error, levien1992output, eschbach1997error, kite2000modeling} can be seen as weighted average of first-order $\SD$ schemes. A general weighted average of that kind -- a $1$st-order  weighted $\SD$ quantizer -- for a line-by-line approach can be characterized by  a set of direction  $\bb d_{i,j}$ with $i\in \{0,\dots, p\}$ and $j\in \{-s,\dots,\ell\}$ for some $\ell, s, p \in \N$ and  a weight  matrix $\W \in \R^{(\ell+s+1) \times (p+1)}$  given by 
\begin{equation}\label{weight-m}
    \W= \begingroup 
\setlength\arraycolsep{2pt}
\begin{pmatrix}
     0 &\cdots &0 & w_{0,1}&  \cdots& w_{0,\ell}\\
     \\
    w_{1,-s}& \cdots & w_{1,0}& w_{1,1}&\cdots & w_{1,\ell}  \\
    \vdots& \ddots & \vdots & \vdots &  \ddots & \vdots\\
    w_{p,-s}& \cdots& w_{p,0}& w_{p,1}& \cdots&  w_{p,\ell}
    \end{pmatrix},
\endgroup \quad \text{with} \quad \quad 
     \\
   \sum\limits_{i=0}^{p}\sum\limits_{j=-s}^{\ell}w_{i,j}=1.  
\end{equation}
Then for a given sample pixel sequence $\p=\{\p_{\bm n}\}_{\bm n \in \N^2}$,  the halftoned pixels are obtained by the following recurrence relation 
\begin{align}
 \label{1st-order-QS-weih_1}
  \sum\limits_{i=0}^{p}\sum\limits_{j=-s}^{\ell}w_{i,j}\left(\Delta_{\bb d_{i,j}} v\right)_{\bm n} &=y_{\bm n}- q_{\bm n}\\[-10pt]
 q_{\bm n}&= \sign\Big( v_{\bm n}- \sum\limits_{i=0}^{p}\sum\limits_{j=-s}^{\ell}w_{i,j}\left(\Delta_{\bb d_{i,j}} v\right)_{\bm n} + y_{\bm n}\Big).  \label{1st-order-QS-weih_2}
 \end{align}
For a visual illustration, see Figure~\ref{fig:my_label}.

Some examples of weight matrices for popular $1$st-order  weighted $\SD$ quantizers include
\vspace{-1mm}
\begin{enumerate}





\item[1)] The Floyd–Steinberg scheme has the weight matrix 
\vspace{-1mm}

\begin{equation*}
    \W_{F\text{-}S}= \left(\begin{matrix}
     0 & \bm{ 0} & \frac{7}{16}\\[1pt]
     \frac{3}{16} & \frac{5}{16}& \frac{1}{16}    \end{matrix}\right).
\end{equation*}
Here and below the element $w_{0,0}$ on the weight matrix is indicated by zero in bold showing the number of negative directions are included.

\item[2)] one of the Shiau-Fan schemes \cite{Shiau1996}, which were introduced as improvements of the Floyd–Steinberg algorithm  corresponds to the weight matrix 

\vspace{-1mm}

\begin{equation*}
     \W_{Sh\text{-}Fan}= \left(\begin{matrix}
 0&  0 & 0 & \bm{ 0} & \frac{8}{16}\\[1pt]
 \frac{1}{16} &\frac{1}{16}     & \frac{2}{16} & \frac{4}{16}& 0   \end{matrix}\right).
 \end{equation*}
\item[3)] The 12-element Jarvis-Judice-Ninke scheme well-known as an edge enhancement technique \cite{Jarvis1976}  corresponds to a matrix $W\in \mathbb{R}^{3\times 5}$.

\end{enumerate}

\begin{figure}
    \centering
    \includegraphics[width=0.55\textwidth]{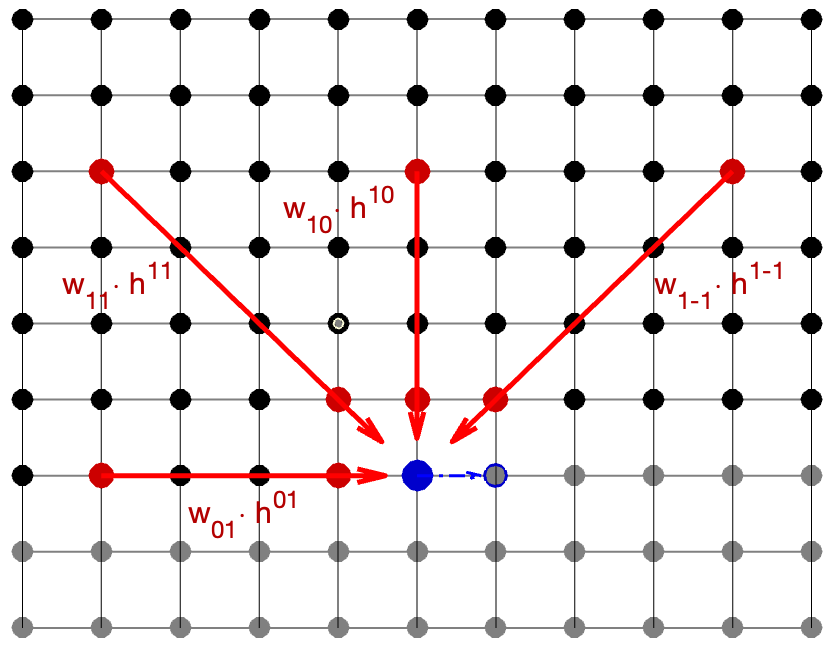}
    \caption{  Visualization of a weighted $\SD$ scheme of second order with weight matrix $\W\in \R^{3\times 2}$ and a  ($2$-sparse) filter $h\in \R^1$. The red dots indicate elements of $\vv$ used in the current quantization step to define the $\vv_{\bm n}$  indicated by a blue dot. Black dots denote already quantized elements, grey dots those not yet quantized. The subsequent element to be quantized is indicated by a blue circle.}
    \label{fig:higher-order-WSD}
\end{figure}

 
 It is remarkable that despite  the superior performance of higher-order $\SD$ scheme in one dimension, most state-of-the-art methods for error diffusion correspond to first order $\SD$ schemes. A possible explanation is related to the amplitude limitations discussed above, which are also confirmed in experiments: a direct application of weighted higher order $\SD$ can lead to severe artifacts in case of black or white regions in the image \cite{Krahmer2023-siam}. The strategy of \cite{Krahmer2023-siam} to overcome this issue is based on a minimal, basically invisible rescaling of the image in combination with a set of feedback filters $h^{i,j}$ adapting to large amplitudes -- one for each direction $\bb d_{i,j}$, see Figure~\ref{fig:higher-order-WSD} for an illustration.
 This gives rise to weighted $\SD$ schemes for image halftoning, which, to the best of our knowledge,  are the first error diffusion schemes that can be interpreted as weighted higher-order $\SD$ schemes.

More precisely, the iterative algorithm reads as follows  
\begin{align}\label{rth-order-sim-QR-filter_1}
  \vv_{\bm n}-\sum\limits_{i \, j}w_{i,j}\left(h^{i,j}*_{\bb d_{i,j}} \vv\right)_{\bm n} &=\p_{\bm n}- \q_{\bm n}\\
 \q_{\bm n}&= \sign\Big(   \sum\limits_{i \, j} w_{i,j}\left(h^{i,j}*_{\bb d_{i,j}}\vv\right)_{\bm n} + \p_{\bm n}\Big), \label{rth-order-sim-QR-filter_2}
 \end{align}
 where the weight matrix $\W \in \R^{(\ell+s+1) \times (p+1)}$ is analogous to  \eqref{weight-m} and the single-index filters $h^{i,j}=\lbc h^{i,j}_n\rbc_{n\in \Z}\in \ell^1(\Z)$. If all filters $h^{i,j}$ fulfill the condition $$\delta^0-h^{i,j}= \Delta^r g^{i,j}$$
 for sequences $g^{i,j}\in \ell_1(\Z)$ with  $g^{i,j}_n=0$ for $n<0$,  we speak of a quantizer of  order $r$.  Moreover, the linear dependence on $v$ in the equations 
 \eqref{rth-order-sim-QR-filter_1}-\eqref{rth-order-sim-QR-filter_2} can be described by a single  extended weight matrix incorporating the effects of both $h$ and $\W$, which is illustrated by several examples below.

Some second-order feedback filters that have proven particularly useful are ${h^{2}:=[0, \; 3/2, \; 0, \; -1/2]}$ and ${h^{3}=[0, \; 4/3, \; 0, \; 0, \; -1/3]}$; they have been proposed in \cite{Gunturk2003, Krahmerthesis, deift2011optimal}. With these filters, one can construct, for instance, the following weighted $\SD$-schemes of the second order. 

\begin{enumerate}
\item[1)]   Simple averageschemes of second order use the weight matrix
\begin{equation*}
    \W_{1/2}= \left(\begin{matrix}
      0 & \frac{1}{2}\\
    
      \frac{1}{2}& 0  \end{matrix}\right), 
\end{equation*}

combined with second-order filters $h^{0,1}\!=\!h^2$  and $h^{1,0}\!=\!h^3$ or $h^{0,1}\!=\!h^{1,0}\!=\!h^3$, respectively, we obtain the extended weight matrices 
\begin{equation*}
{\widetilde{\W}}^{2nd}_{A23}= \left(\begin{matrix}
     \bm 1 & -\frac{3}{4} & 0 &  \frac{1}{4}\\
     -\frac{4}{6}&0& 0&0 \\
     0 & 0& 0&0  \\[-2pt]
     0 & 0& 0&0 \\[-1pt]
\frac{1}{6} &0& 0&0 \\
      \end{matrix}\right),  
      \quad \quad 
    {\widetilde{\W}}^{2nd}_{A33}=  \left(\begin{matrix}
     \bm 1 & -\frac{4}{6} & 0 & 0& \frac{1}{6}\\
     -\frac{4}{6}&0& 0&0 &0\\
     0 & 0& 0&0 &0 \\[-2pt]
     0 & 0& 0&0 &0\\[-1pt]
\frac{1}{6} &0& 0&0 &0\\
      \end{matrix}\right), 
\end{equation*}

\item[2)] The  Floyd–Steinberg schemes of second order 
with all four feedback filters $h^{i,j}$ chosen to be  $h^3$  yields the extended weight matrix
\begin{equation*}
{\widetilde{\W}}^{2nd}_{F\text{-}S\text{-}33}=  \left(\begin{matrix}
    0 & 0&  0 & 0&   \bm 1 & -\frac{28}{48} & 0 & 0& \frac{7}{48}\\[2pt]
   0 & 0&  0 & -\frac{12}{48}&    -\frac{20}{48}&-\frac{4}{48} & 0&0 &0\\
   0 & 0&  0 & 0&    0 & 0& 0&0 &0 \\[-2pt]
   0 & 0&  0 & 0&    0 & 0& 0&0 &0\\[-2pt]
 \frac{3}{48} & 0&  0 & 0& \frac{5}{48} &0& 0&0 &\frac{1}{48} \\
      \end{matrix}\right).
\end{equation*}
\end{enumerate}

In the similar fashion, one can also design the second-order Shiau-Fan and Jarvis-Judice-Ninke schemes. 

The halftoned images obtained by some of these second-order $\SD$ schemes in combination with a minimal rescaling indeed led to halftoned images with better similarity indices than state-of-the-art approaches \cite{Krahmer2023-siam}. Also mixed-order schemes with third order contributions could be demonstrated to yield an additional minimal improvement. 

Furthermore, for the bandlimited model, one can also assess the theoretical performance.  The following theorem provides an estimate for the halftoning error of weighted $\SD$ schemes in terms of the oversampling rate.

\begin{thm}\label{thm-2d-error-high-order-WSD}  \cite{Krahmer2023-siam}
Consider a bandlimited function $w\in  \B^b$ sampled on the lattice $ \frac{1}{\lam}\,\N^2$ with the oversampling rate $\lam>1$. If a weighted $\SD$ scheme \eqref{rth-order-sim-QR-filter_1}-\eqref{rth-order-sim-QR-filter_2} used for construction of $w$'s 1-bit samples $\q\!\in\!\{-1,1\}^{\N^2}$ is of order $r$, then the corresponding quantized representative $w_{ q}$ satisfies 
\begin{equation}\label{error-bound-high-order}
    \nofty{w_{\lam} - w_q} \le  \frac{1}{\lam^r} \,    \nofty{v} \Big( C_{\W}\cdot C \cdot \|\nabla^r \Phi\|_{1,2}  + \mathcal{O}(\lam^{-1})\Big),
\end{equation}
where $C>0$ is a constant independent of $\W$, $\Phi$ is a Schwartz function of the low-pass type,  the weight constant  $C_{\W}$ depends the feedback filters $h^{i,j}$. 
   
\end{thm}

It should be noted, however, that the $\ell_\infty$ error captured by this theorem does not directly correspond to visual similarity. In particular, optimizing the filters for this metric has been demonstrated to yield suboptimal results in terms of similarity indices \cite{Krahmer2023-siam}. Still, the performance of the higher order $\SD$ schemes show that the benefits of higher order can outweigh the effects of model mismatch, so the mathematical theory for $\SD$ can provide a guiding intuition for digital halftoning.
\section{Outlook and Open Problems}

While the aforementioned results already give significant insights about halftoning, many problems remain to be answered. 
On the one hand, for the attraction-repulsion model, it would be interesting to investigate how the discretization error affects the reconstruction quality. This would make this class of methods directly applicable to the discrete setting and allow for a quantitative comparison to error diffusion approaches.

On the other hand, the theoretical underpinning of the error diffusion schemes via weighted  $\SD$ schemes is currently available only on $\R^2$. Higher dimension, in particular 3D printing and video halftoning, pose additional challenges. For video, the role of the dimensions is quite different. In particular, a challenge of video halftoning is to avoid flicker effects \cite{rehman2011alleviating,schmaltz2012video, rehman2010flicker}.  We expect that to address this issue,  one needs to design adaptive error diffusion methods, allowing only for a change of pixels in the halftoned version when they are changing in the video.  

In addition, halftoning problems are also interesting for geometries beyond Euclidean space. For manifolds and also for graph domains, initial steps towards understanding $\SD$ quantization are available, but currently only for very special cases \cite{graf2023one, krahmer2023quantization}, and the implications for halftoning remain to be discussed.  

Lastly, a recent paper  \cite{lyu2023sigma} has devised two-bit schemes for the related problem of image quantization. We find it interesting question whether these schemes can be generalized to one-bit alphabets and, if yes,  how they can be used for halftoning.


\end{document}